 \documentclass[manuscript]{aastex62}
\graphicspath{{./}{figures/}}

\received{}
\revised{}
\accepted{}
\submitjournal{ApJ}

\shorttitle{Flat energy spectra}
\shortauthors{Perri et al.}

\usepackage{soul,xcolor}

\newcommand{\noopsort}[1]{}

\begin{document}

\title{Interpretation of flat energy spectra upstream of fast interplanetary shocks}

\correspondingauthor{Silvia Perri}
\email{silvia.perri@fis.unical.it}

\author[0000-0002-8399-3268]{Silvia Perri}
\affil{Dipartimento di Fisica, University of Calabria, Rende, Italy}

\author{Giuseppe Prete}
\affil{Dipartimento di Fisica, University of Calabria, Rende, Italy}

\author{Gaetano Zimbardo}
\affil{Dipartimento di Fisica, University of Calabria, Rende, Italy}

\author{Domenico Trotta}
\affil{The Blackett Laboratory Imperial College London, London SW7 2AZ, UK}

\author{Lynn B. Wilson III}
\affil{Heliophysics Science Division, NASA Goddard Space Flight Center, Greenbelt, MD 20771, USA}

\author[0000-0002-3176-8704]{David Lario}
\affil{Heliophysics Science Division, NASA Goddard Space Flight Center, Greenbelt, MD 20771, USA}

\author{Sergio Servidio}
\affil{Dipartimento di Fisica, University of Calabria, Rende, Italy}

\author{Francesco Valentini}
\affil{Dipartimento di Fisica, University of Calabria, Rende, Italy}

\author{Joe Giacalone}
\affil{Lunar and Planetary Laboratory, University of Arizona, Tucson, USA}

\begin{abstract}

Interplanetary shocks are large-scale heliospheric structures often caused by eruptive phenomena at the Sun, and represent one of the main sources of energetic particles. Several interplanetary shock crossings by spacecraft at $1$ AU have revealed enhanced energetic-ion fluxes that extend far upstream of the shock. Surprisingly, in some shock events, ion fluxes with energies between $100$ keV and about $2$ MeV acquire similar values (which we refer to as ``overlapped'' fluxes), corresponding to flat energy spectra in that range. In contrast, closer to the shock, the fluxes are observed to depend on energy. In this work, we analyze three interplanetary shock-related energetic particle events observed by the Advanced Composition Explorer spacecraft where flat ion energy spectra were observed upstream of the shock. We interpret these observations via a velocity filter mechanism for particles in a given energy range. This reveals that low energy particles tend to be confined to the shock front and cannot easily propagate upstream, while high energy particles can. The velocity filter mechanism has been corroborated from observations of particle flux anisotropy by the Solid-State Telescope of Wind/3DP.

\end{abstract}



\section{Introduction} 
\label{sec:intro}

Collisionless shock waves are observed to be one of the main sources of energetic particles and cosmic rays in astrophysical environments. Efficient particle acceleration at shocks can result when particles remain confined near the shock, either by scattering in turbulent magnetic fields, or as a result of the geometry, and gain energy due to the compression in the plasma velocity at the shock front \citep[e.g.,][]{Drury1983}. Shock acceleration is supported by in-situ measurements in the heliosphere, where energetic particle fluxes are clearly peaking at the time of shock passage \citep[][]{Giacalone2012}, and by remote observations of supernova remnants \citep[e.g.,][]{Bamba03,Morlino10,Reynoso13}. On the other hand, the quantitative agreement with the predictions of diffusive shock acceleration (DSA), like spectral slope, acceleration times, and maximum energies, remains elusive \citep{Lagage83,Giacalone2012,Vainio14}. For instance, to reach an energy of $\sim 10^{15}$ eV (i.e., the so-called knee of the cosmic ray energy spectrum), a substantial amplification of the pre-existing (upstream) magnetic field by means of self-generated turbulence is required \citep[e.g.,][]{Drury1983,Blasi13,Amato14}. In a similar way, the particle mean free paths in the heliosphere are estimated to be in the range of 0.01--1.0 AU, but this is far too large to explain the rapid acceleration of particles by interplanetary (IP) shocks \citep{Perri15}. Thus, turbulence enhancement, related to the backstreaming of energetic particles upstream of IP shocks, is required \citep[e.g.,][]{Ng03,Afanasiev18}. Similar electromagnetic fluctuation amplification is required to explain the efficient acceleration of solar energetic particles (SEPs) by CME-driven shocks \citep[e.g.,][]{Lee12}. 

In collisionless plasmas, because of particle reflection at supercritical shocks \citep{burgess_book}, shock properties also depend on the (acute) angle, $\theta_{\rm Bn}$, between the average upstream magnetic field, $\bf B$, and the unit-normal vector to the shock. If $\theta_{\rm Bn}< 45^{\circ}$ the shock is termed quasi-parallel and reflected ions can efficiently propagate upstream, thereby forming the ion foreshock. Here, reflected particles form a beam in velocity space which can excite ion cyclotron, Alfv\'en waves, or fast-magnetosonic, and whistlers, making the ion foreshock a very turbulent region, as often observed by spacecraft in the terrestrial ion foreshock \citep[e.g.,][]{Schwartz91,Perri09,Wilson16a,Karimabadi14}. These disturbances can give the required enhanced level of magnetic fluctuations to efficiently scatter both electrons \citep{Wilson16b} and ions \citep{Turner18,Wilson13}, thereby trapping them near the shock for further acceleration to high energies according to the mechanism of DSA.
Conversely, if $\theta_{\rm Bn}> 45^{\circ}$ the shock is termed quasi-perpendicular and most reflected ions re-enter the shock, contributing to enhanced level of fluctuations downstream, as usually observed \citep{Greenstadt75}. In the absence of rapid scattering upstream, DSA can be slow, as suggested by some observations \citep{Reynoso13} and numerical simulations \citep[e.g.,][]{Caprioli2015,Sundberg2016,Trotta20,Preisser20}.  
However, it should be noticed that enhanced levels of magnetic fluctuations upstream of quasi-parallel IP shocks are not always observed \citep{BlancoCano16}; for instance, in a recent study \citet{Zim20} found that the magnetic power at the scales corresponding to energetic particle resonance was approximately constant from upstream to downstream for three IP shocks with different values of $\theta_{\rm Bn}$. Therefore, the conditions under which self-generated turbulence is actually found need to be better understood. 

Observations of fluxes of supra-thermal particles can be highly influenced by the upstream transport conditions, which are established by the level of magnetic field fluctuations. For example, \citet{Lario22} investigated the formation of an anisotropic field-aligned beam of protons upstream of an oblique shock with energies $\le 30$ keV together with a population of protons at higher energy propagating at small pitch-angle. The unusually long duration (and therefore spatial extent) of the field-aligned beam was interpreted as due to the absence of magnetic field fluctuations over a large distance upstream of the shock wave --- a scenario where efficient scattering is not favoured.


Another fundamental phenomenon is the formation of a precursor of energetic particles in the upstream: this precursor is distinct from the ion foreshock (which is formed mostly by reflected thermal particles), or from the fast/magnetosonic whistler wave precursor \citep{Wilson17}, and is due to those energetic particles satisfying 
\begin{equation}
v\mu>V_1^{sh} \sec(\theta_{\rm Bn}),
\label{eq:filter}
\end{equation}
i.e., those particles that have velocity parallel to the local magnetic field (being $v$ the particle speed and $\mu$ the cosine of the pitch angle) larger than the speed of the intersection point of a field line with the shock front along the direction of the shock front \citep[e.g.,][]{LeRoux12,lario19}. Notice that $V_1^{sh}$ represents the upstream solar wind speed in the frame of reference of the shock.

This velocity filter can lead to a spectral flattening upstream of the shock, since only higher energy particles, can easily propagate back upstream, while lower energy particles tend to be confined within the shock region, reducing their fluxes far upstream. Nonetheless, the presence of turbulent fluctuations causes pitch angle scattering and a meandering of magnetic field lines, which can influence the velocity filter effect on particles.

Such spectral flattening has been predicted in earlier works by \citet{Ng03}, as a consequence of the differential growth of Alfv\'en waves upstream of the shock, amplified by high energy ions streaming away from the shock. This creates a sort of delay in the lower energy particle transport upstream, favouring the formation of a flat energy spectrum. That model, however, suffers the limitation of predicting very short intervals for spectral flattening. More recently, overlapping particle fluxes have been modelled by \citet{Prinsloo19}, and explained as a balance between shock acceleration and adiabatic cooling; however, at $1$ AU they obtained a very narrow (in energy) spectral flattening. Thus, mismatches with observations are found. 

In this paper, we analyze three IP shock crossings and associated energetic ion intensities observed by the Advanced Composition Explorer (ACE) and the Wind spacecraft on 2005 May 15, on 2012 July 14, and on 2003 November 4. These events were chosen because continuous data measurements from various instruments are available, no particular IP disturbances affect energetic-ion fluxes, and the IP shocks have similar compression ratios. 
All these events are characterized by energetic particle profiles that are overlapping far upstream and up to about 1-3 hours before the shock passage (i.e., the upstream energy spectrum is flat for a broad range of particle energies). Closer to the shocks, the fluxes at each energy separate and reach values more typical of those predicted by DSA theory. This peculiar feature of upstream overlapping particle fluxes was first reported by \citet{Lario18}, looking at particles upstream of IP shocks using ACE measurements at energies between approximately 50 keV and 4 MeV. They noted that the region of spectral flattening for these energetic protons happens prior of the shock arrival, as the energy spectrum steepens close to the shock, and the observed spectral shapes are closely related to the particle transport processes from the evolving shock to the observer. 

Here, we analyze energetic particle fluxes and magnetic fluctuations close to and far upstream of the shocks. We interpret observations of flat upstream energy spectra in terms of the velocity filter mechanism described above. Importantly, we derive how particle fluxes are modified from such a condition for different values of the angle $\theta_{\rm Bn}$ and of the upstream plasma speed as measured in the shock frame. 

The paper is organized as follows: Section \ref{sec:data} contains the data description and the data analysis; Section \ref{sec:discussion} presents the interpretation of the data according to the velocity filter condition; in Section \ref{sec:conclusions} concluding remarks are given.

\section{Data collection and analysis} 
\label{sec:data}

\subsection{Shock crossings}
We first analyze the shock event of 2005 May 15, which exhibits overlapping energetic particle intensity profiles over a large distance upstream of the shock crossing. Energetic particle fluxes separate at about $50$ min before the shock crossing and downstream of the shock, thus resulting in a steeper energy spectrum \citep{Lario18}. The solar origin of this relatively strong event was associated with a halo, fast (plane-of-sky speed $\sim$1689 km s$^{-1}$) coronal mass ejection (CME) observed at 17:12~UT on 2005 May 13, as reported in the Coordinated Data Analysis Workshops (CDAW) Data Center SOHO LASCO CME catalog\footnote{available at cdaw.gsfc.nasa.gov/CME$\_$list/} and temporary associated with a
M8.0 GOES class X-ray flare from the NOAA active region 10759 at N12E12 with onset at 16:13~UT on 2005 May 13 \citep{Lario18}. The transit time for the shock to travel from the Sun to 1~AU was about 2038 minutes corresponding to an average transit speed of $\sim$$1223$ km s$^{-1}$ \citep[see Table 1 in][]{Lario18}. Using data from GOES-11, \citet{Lario18} show that energetic particles peak at the shock even in the high energy channel of $32.5$--$56.4$ MeV, confirming that it is a strong particle accelerator.  

\begin{figure}[t]
    \centering
    \includegraphics[scale=0.12]{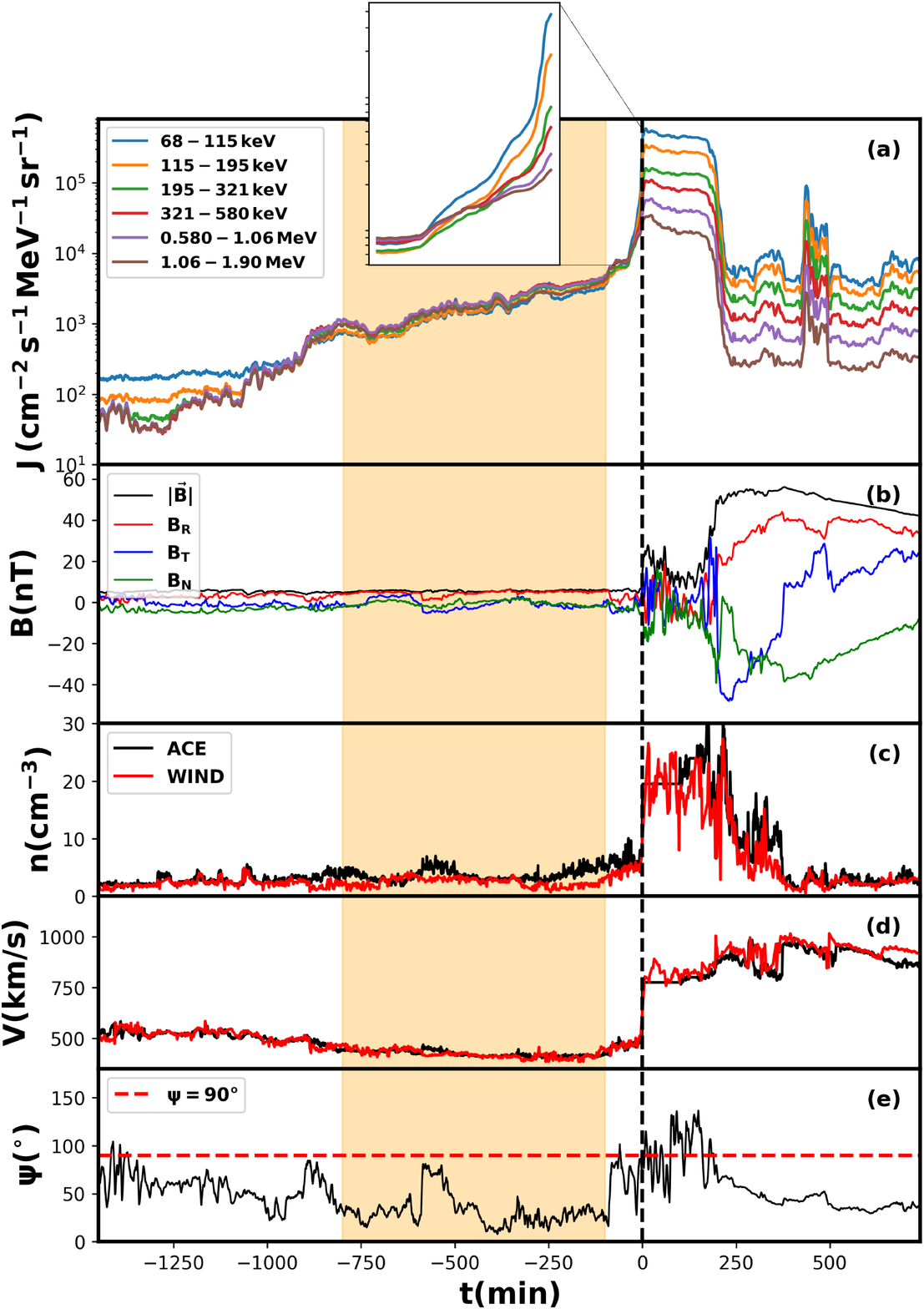}
    \caption{Overview of the shock crossing by the ACE spacecraft on 2005 May 15. From top to bottom: particle fluxes from the EPAM-LEMS120 instrument in the energy range $68$-$1890$ keV measured at $12$ s cadence, the $60$ s resolution magnetic field components in the Radial-Tangential-Normal (RTN) reference frame along with the magnetic field magnitude, the proton density and the solar wind bulk speed both from ACE (at $64$ s resolution) and Wind (at $92$ s resolution), and the angle between the radial direction and the mean field direction computed over a running window of $5$ min. The vertical dashed line indicates the shock position, the yellow shaded area shows the region where energetic particle fluxes overlap indicating a period with a relatively flat energy spectrum.}
    \label{fig:15052005}
\end{figure}

Figure \ref{fig:15052005} shows an overview of the 2005 May 15 event. From top to bottom we report (a) the $12$ s resolution particles fluxes detected by the Electron, Proton, and Alpha Monitor (EPAM) instrument \citep{Gold98} onboard ACE from the LEMS120 sensor in the energy range $68$ keV-$1.90$ MeV; (b) the magnetic field components in  Radial-Tangential-Normal (RTN) coordinates as sampled by the MAG instrument \citep{Smith98} onboard ACE at a time cadence of $60$ s; (c) and (d) the plasma proton density and the solar wind bulk speed, respectively, as detected by the Solar Wind Electron, Proton, and Alpha Monitor (SWEPAM) experiment \citep{McComas98} onboard ACE (black line) at $64$ s resolution. Because of a data gap in the ACE time series around the shock crossing, we add the proton solar wind density measured by the Solar Wind Experiment (SWE) \citep{Ogilvie95} on board Wind at $92$ s resolution (red line). The panel (e) shows the angle $\psi$ between the mean field and the radial direction \citep[e.g.,][]{Sonnerup1969}. The latter gives us insights into the magnetic connectivity of the observer with the shock surface (the horizontal dashed red line indicates a $90^{\circ}$ angle). The angle $\psi$ has been computed using 5-minutes running average windows.

The vertical black dashed line indicates the shock crossing time. We display a time window extending towards $\simeq 24$ hours before the ACE shock crossing. For a period of about $50$ min up to $16$ hours prior to the shock crossing (indicated by a shaded region in Figure \ref{fig:15052005}) the particle fluxes at the displayed energies overlap. For earlier times (i.e. more than $17$ hours prior to the shock crossing), particle fluxes at different energies are well separated. It can be seen in panel (b) that within the region upstream where the energetic particle fluxes overlap, the magnetic field tends to be approximately radial (see also panel (e) in Figure~\ref{fig:15052005} where $\psi<30^{\circ}$ for most of the time interval indicated by the shaded region), while it largely deviates from the radial direction close to the shock. This suggests that for the entire period when the upstream flat spectrum was observed, the spacecraft was well connected to the shock front.
However, the fluxes remain overlapped even closer to the shock when the field starts changing direction. Indeed, fluxes separate only very close to the shock ($-40$ min), while the magnetic field deviates from the radial direction at about $-85$ min.

In the downstream region, $\sim$3 hours after the shock crossing, a sudden decrease of about two orders of magnitude in the particle fluxes is observed, which is due to the arrival of the magnetic obstacle of the CME that generated the SEP event \citep{Lario18,Perri22}.

\begin{table}
\caption{Main parameters for the three shock crossings analyzed, as deduced from ACE and Wind measurements (see text for details). The $\theta_{Bn}$ estimates using both the minimum variance (MVA) and the magnetic coplanarity (MC) methods are reported, the Alfv\'enic and the sonic mach numbers, the compression ratio of the shock, the plasma beta, and the shock speed in the satellite frame. Associated errors are also reported.}\label{table1}
\centering
\begin{tabular}{ccccccccc}
\hline
 date &  time (ACE) &   $\theta_{Bn} (^{\circ})$-MVA  & $\theta_{Bn} (^{\circ})$-MC &  $M_{A}$ & $M_{S}$ & $r$ & $\beta_{plasma}$ & $V_{sh}^{s/c}({\rm km/s})$ \\
\hline

$15/05/05$ & 02:05 &  $78.7 \pm 12.9 $ & $51.3 \pm 13.6 $ & $11.6$ &  $ 17.5$ &  $3.0 \pm 0.6$ & $1.57$ & $926.9$ \\

$14/07/12$ & 17:26 &  $68.6 \pm 22.9 $ & $67.6 \pm 7.3 $ & $42.7$ &  $81.4 $ &  $3.0 \pm 0.5$ & $0.99$ & $710.6$\\

$04/11/03$ & 05:59 &  $43.4 \pm 12.5 $ & $9.9 \pm 5.7 $ & $6.01$ &  $10.5 $ &  $3.6 \pm 1.1 $ & $1.2$ & $803.7$\\

\hline
\end{tabular}
\end{table}


Fundamental shock parameters have been computed and are shown in Table \ref{table1}, although single-spacecraft shock parameter estimation is a process that is known to be subject to many sources of uncertainty \citep{Paschmann2000, Koval2008}. For $\theta_{\rm Bn}$, we used both the Minimum Variance Analysis (MVA) over a 30 minutes time interval before the shock crossing, similarly to what has been done in \citet{vanNes84}, although this technique can be highly unstable because of rapid variations of the magnetic field, and the magnetic coplanarity (MC) method \citep{Colburn1966}. The MC technique with a systematic variation of upstream and downstream averaging windows from $\sim$ 2 to 30 minutes \citep[as done in][]{Trotta2022b} yields a large spread of values, indicating high levels of upstream/downstream disturbances. The Alfv\'en and the sonic Mach numbers, the compression ratio of the shock, the plasma beta $\beta=nk_BT/(B^2/2\mu_0)$ (being $k_B$ the Boltzmann constant and $\mu_0$ the vacuum magnetic permeability), and the shock speed in the spacecraft frame of reference \citep{PerriEA15} are also reported (values averaged over a time window of 30 min). It should be also noted that the parameter estimations assume a stationary, planar, and infinite shock, a hypothesis that is often not satisfied. 
Actually, the shock structure can be further complicated by the interaction with pre-existing structures that modify the local geometry of the shock front (as recently addressed for IP shocks \citep{BlancoCano2019,Giacalone2021} as well as for the Earth's bow shock \citep{Trotta2022a}). Moreover, the shocks analysed here are supercritical (see Table \ref{table1}), indicating that they may be affected by many self-induced spatial and temporal irregularities happening at a variety of scales  \citep[e.g.][]{Kajdic2019,Kajdic2021}.

From Table \ref{table1}, the 2005 May 15 event is a quasi-perpendicular shock, with sonic and Alfv\'enic Mach numbers $M_s\sim M_A \sim 16$, which is able to accelerate ions to tens of MeV at $1$ AU (i.e., at the spacecraft position).


\begin{figure}[t]
    \centering
    \includegraphics[scale=0.65]{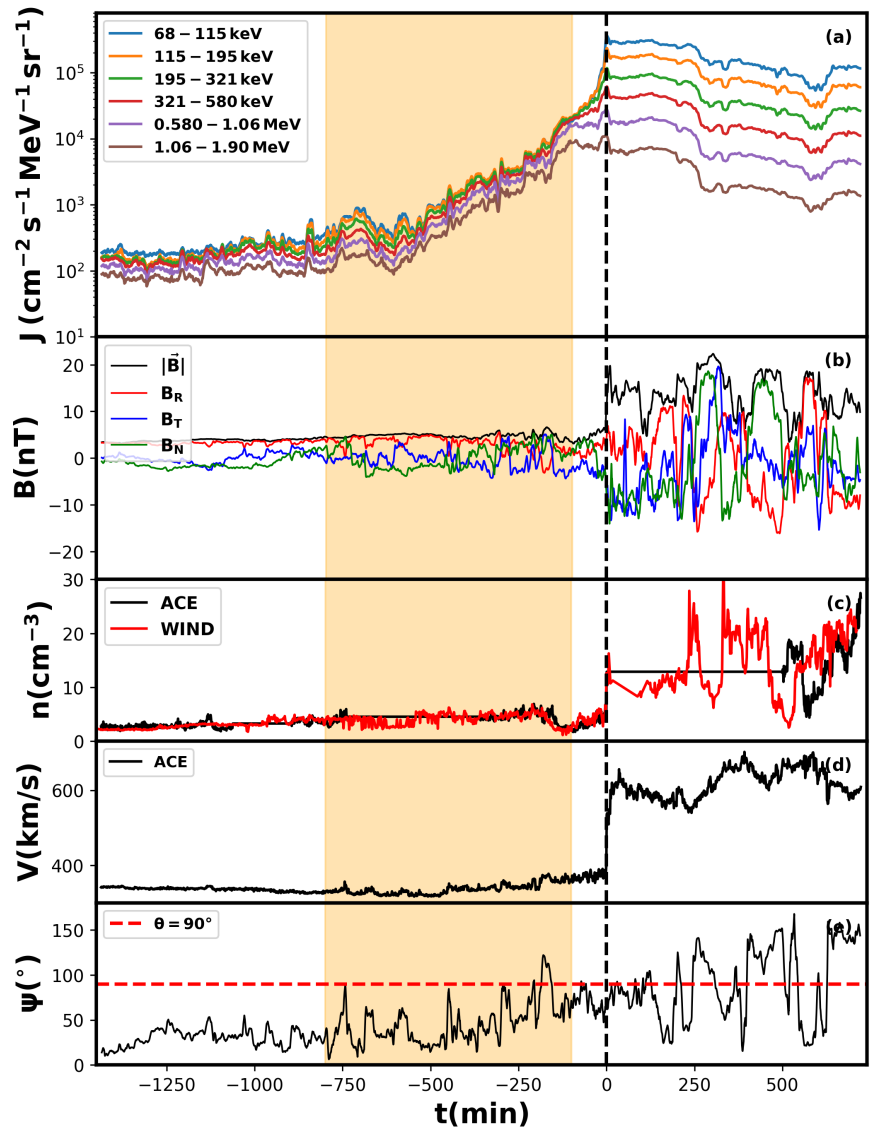}
    \caption{Same as Figure \ref{fig:15052005} but for the shock crossing by the ACE spacecraft on 2012 July 14.}
    \label{fig:14072012}
\end{figure}

\begin{figure}[t]
    \centering
    \includegraphics[scale=0.65]{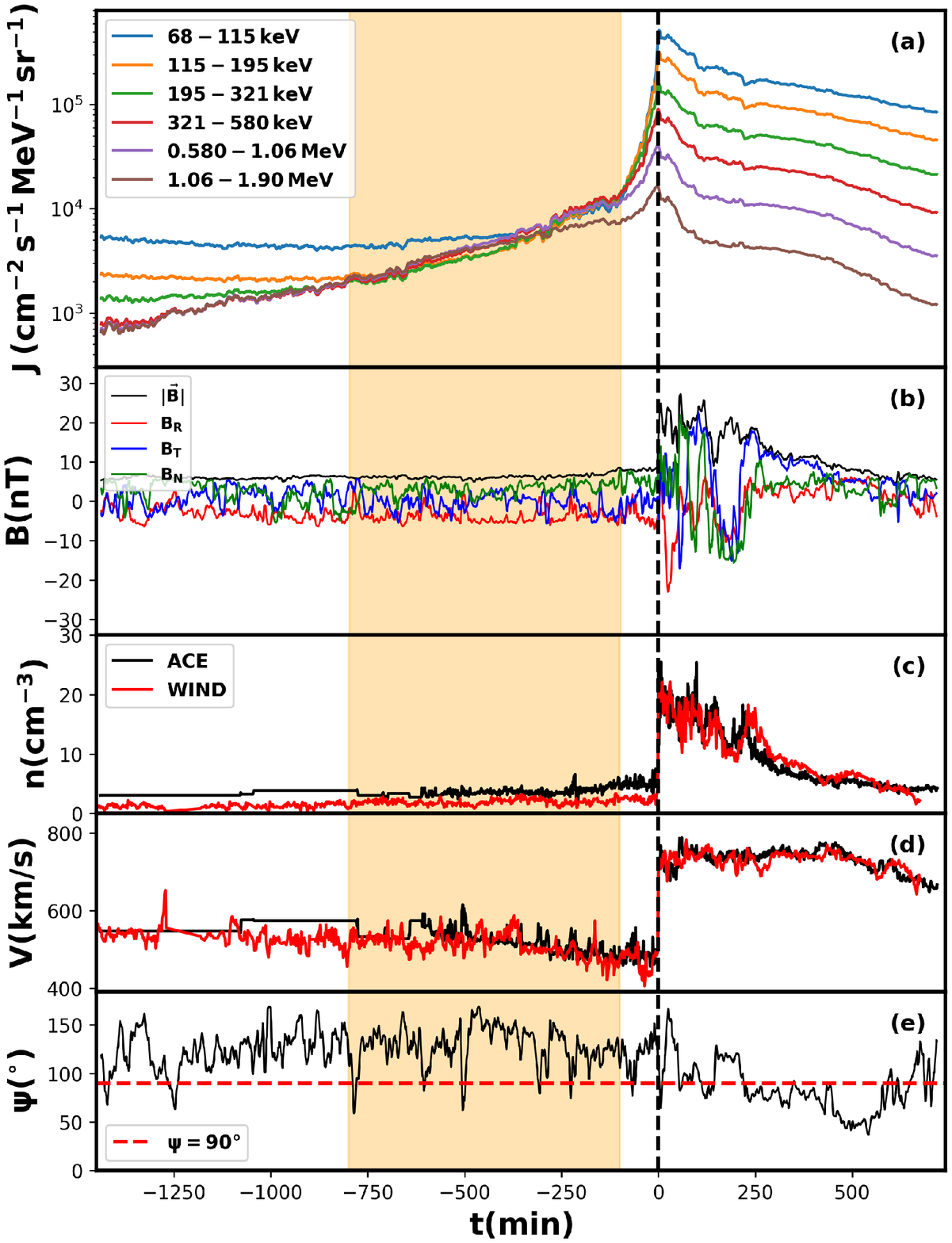}
    \caption{Same as Figure \ref{fig:15052005} but for the shock crossing by the ACE spacecraft on 2003 November 4.}
    \label{fig:04112003}
\end{figure}

The second analyzed event is a quasi-perpendicular shock at 17:26~UT on 2012 July 14. 
The solar origin of this event was associated with a 
fast (plane-of-sky speed 885 km s$^{-1}$) halo CME observed at 16:48~UT on 2012 July 12 as reported by the CDAW SOHO LASCO CME Catalog and temporally associated with 
a X.4 flare
from NOAA Active Region 11520 at S15W01 at 15:37~UT \citep{wijsen22}. 
The transit time of the shock to travel from the Sun to 1~AU is about 2989 minutes, corresponding to an average transit speed of $\sim$ 834 km s$^{-1}$. 
Figure~\ref{fig:14072012} shows data for the event on 2012 July 14 (with the same format as Figure~\ref{fig:15052005}).
This event displayed a flat particle energy spectrum from more than $12$ hrs to about $2$ hrs prior to the shock passage (shaded region in Figure \ref{fig:14072012}). In this case the magnetic field was almost radial ($\psi<50^{\circ}$) but gradually started to change direction about $4.5$ hrs prior to the shock crossing, while the fluxes remained overlapped (at least the ones in the lower energy channels, from $68$ keV to $580$ keV).

The third shock crossing analyzed occurred at 05:59~UT on 2003 November 4 and corresponds to a quasi-parallel shock (see Table \ref{table1}).
The solar origin of this event was associated with 
a fast (plane-of-sky speed 2598 km s$^{-1}$) halo CME at 17:30~UT on 2003 November 2 as reported by the CDAW SOHO LASCO CME Catalog and temporally associated with 
a X8.3 flare at 17:03~UT from NOAA Active Region 10486 at S14W56 \citep{lario05}. 
The transit time of the shock to travel from the Sun to 1~AU is about 2216 minutes, corresponding to an average transit speed of $\sim$1125 km s$^{-1}$. 
Figure \ref{fig:04112003} shows, with the same format as Figure~\ref{fig:15052005}, data for the event on 2003 November 4.
Energetic ion fluxes overlapping for several hours before the shock passage at the spacecraft position within the energy channels between $115$ keV and $1.06$ MeV. The ion fluxes prior to the shaded region in Figure \ref{fig:04112003} were already elevated due to prior SEP events.
We note that the fluxes in the lowest energy channel (i.e., $68$--$115$ keV) and in the highest energy channel ($1.06$--$1.90$ MeV) overlap partly within the shaded box in Figure \ref{fig:04112003}. The magnetic field is oriented sunward in this period, oscillating around $\psi\sim 135^{\circ}$.

\subsection{Magnetic field turbulence}
In order to characterize the environment through which the shocks propagate, we have studied the properties of magnetic field turbulence within both (i) the upstream region where flat energy spectra are detected, and (ii) close to the shock where power-law energy spectra are observed. This has been done by computing the power spectral density (PSD) of the magnetic field fluctuations along each direction in the RTN reference frame, using the Fast Fourier Transform. 

\begin{figure}[t]
    \centering
    \includegraphics[scale=0.65]{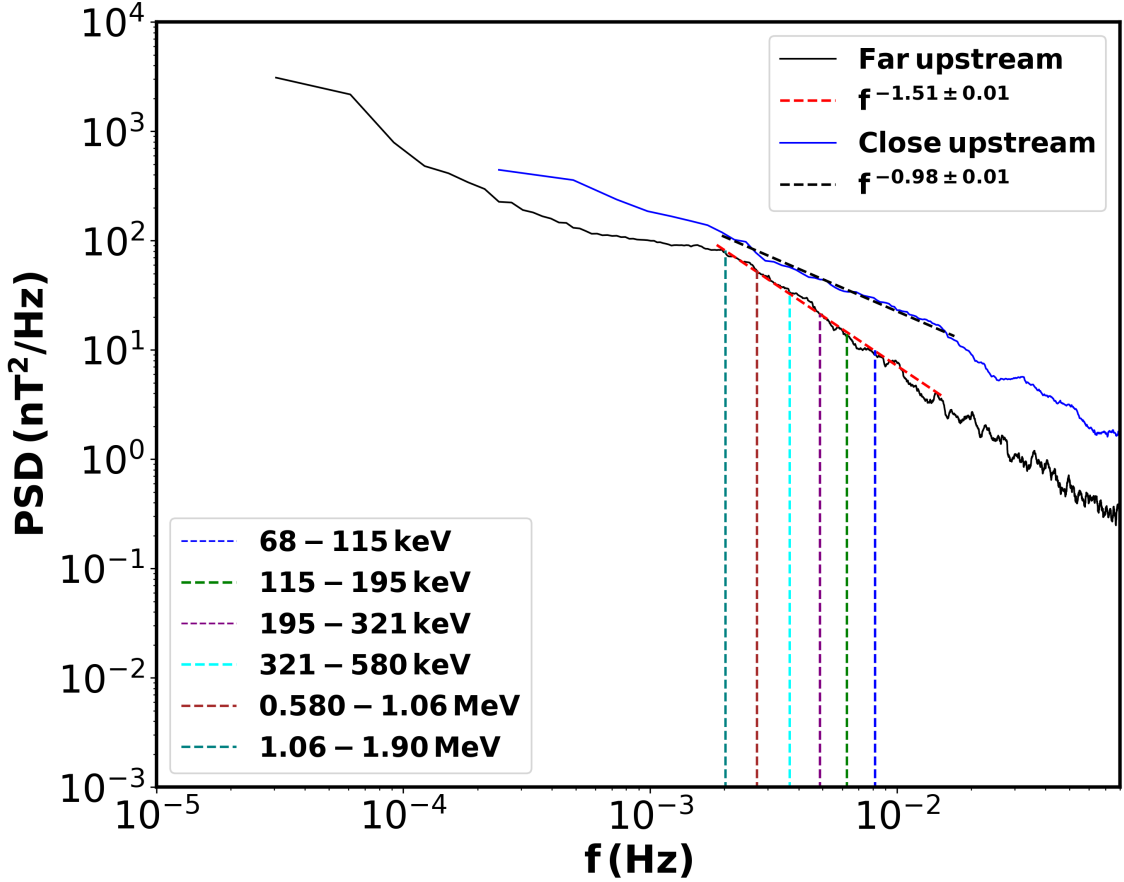}
    \caption{PSD of the magnetic field fluctuations for the 2005 May 15 event, computed in the region far upstream where the fluxes of energetic particles overlap (black line) and in the region close to the shock front upstream (blue line). The vertical dashed lines indicates the frequencies corresponding to the Larmor radius of the energetic particles (see Figure legend). The best power-law fits are also shown by the dashed lines.}
    \label{fig:psd15052005}
\end{figure}

\begin{figure}[t]
    \centering
    \includegraphics[scale=0.65]{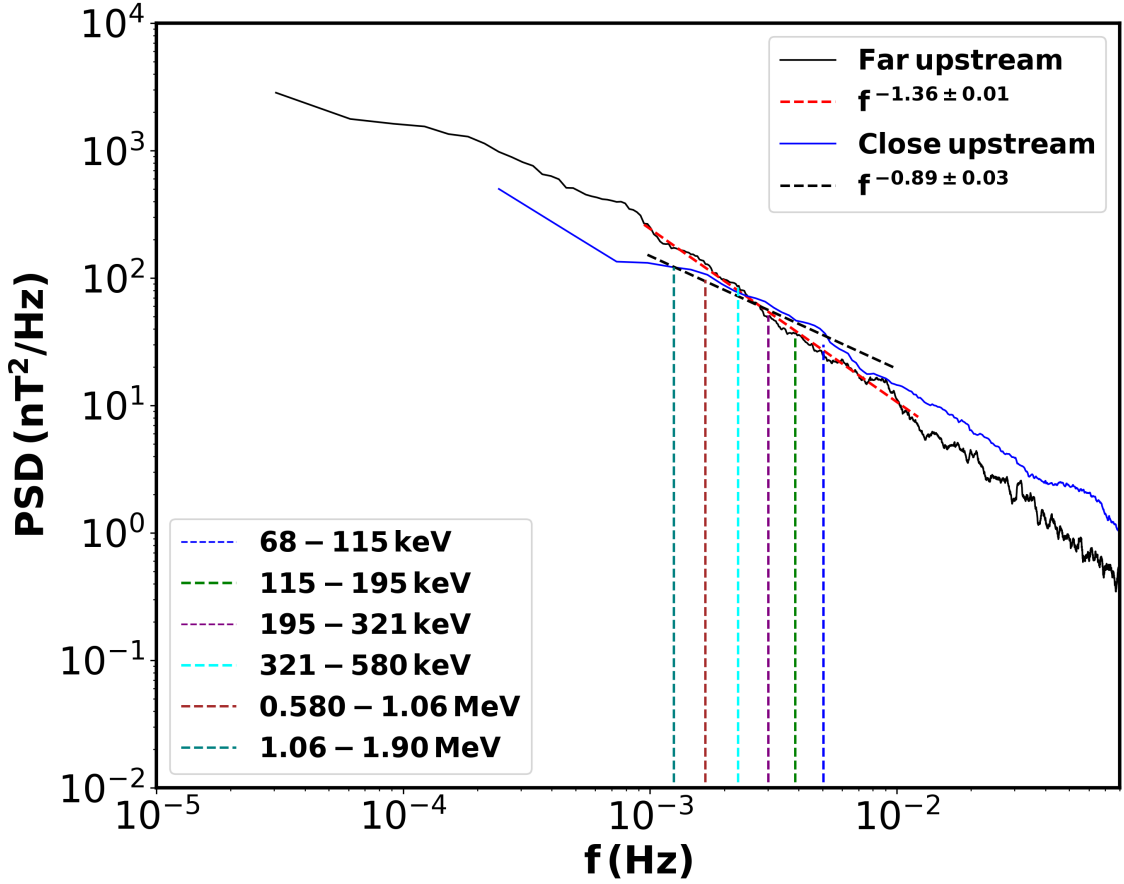}
    \caption{Same as Figure \ref{fig:psd15052005} but for the shock on 2012 July 14.}
    \label{fig:psd14072012}
\end{figure}

\begin{figure}[t]
    \centering
    \includegraphics[scale=0.65]{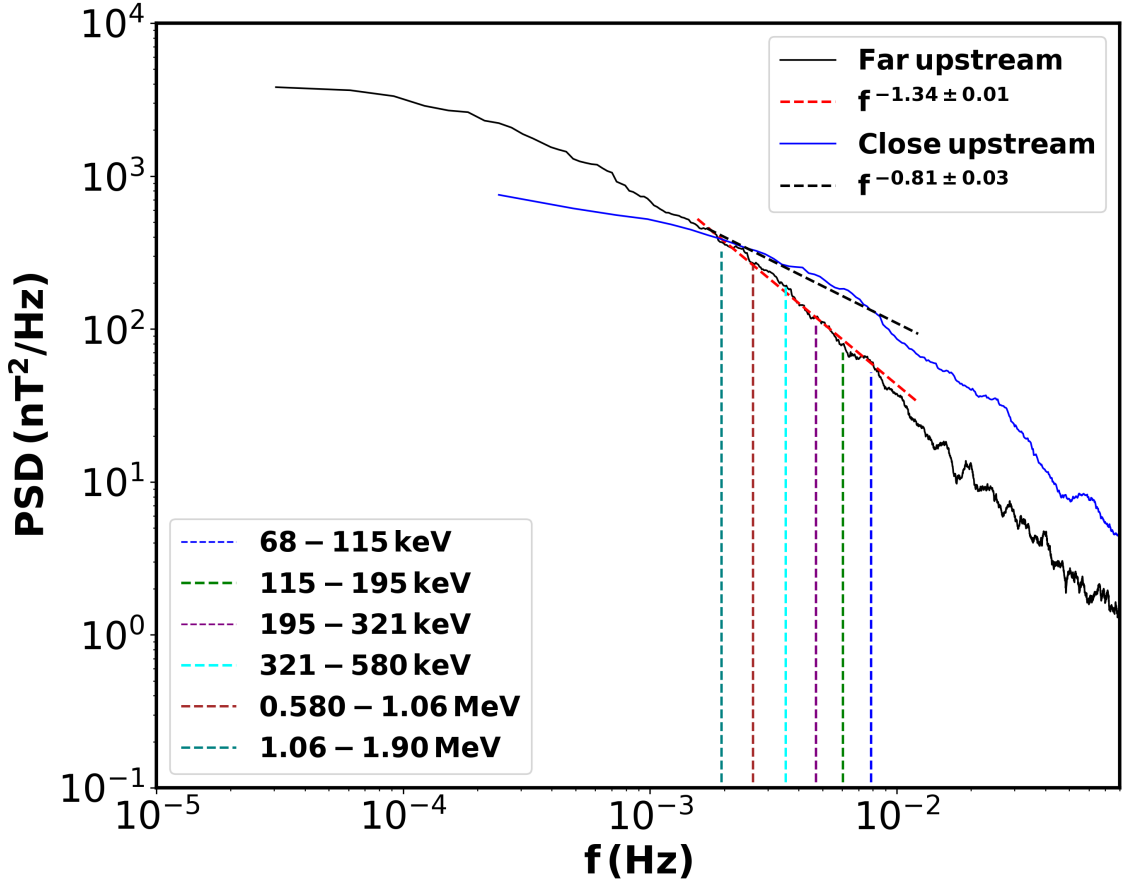}
    \caption{Same as Figure \ref{fig:psd15052005} but for the shock on 2003 November 4.}
    \label{fig:psd04112003}
\end{figure}

Figure \ref{fig:psd15052005} displays the power spectrum of the magnetic field in the two regions described above for the event on 2005 May 15. In the region where the energetic particle fluxes overlap, from -800 min to -100 min from the shock, the spectrum follows a Kolmogorov-like power-law, i.e., ${\rm PSD}(f)\propto f^{-5/3}$, where $f$ is the frequency, indicating ambient solar wind turbulence \citep{Bruno13,Pitna21}. By contrast, in the close upstream region, from -80 min to -10 min, we find ${\rm PSD}(f)\propto f^{-1}$ for the 2005 May 15 event, while in the other two crossings (where we have used the same time interval for PSDs computation) the PSD is bumped at low frequencies (see Figures \ref{fig:psd14072012} and \ref{fig:psd04112003}). In Figure \ref{fig:psd15052005} the best power-law fits are displayed with the black and the red dashed lines. The values found for the slopes, along with their associated errors, are also reported in the legend. Notice that we have computed the power-law fits within a range of frequencies corresponding to the Larmor radius of the energetic particles, namely, $f_{E}=V_{\rm up}/(2\pi \rho_{E})$ where $V_{\rm up}$ is the proton bulk speed in the relevant upstream region in the spacecraft frame and $\rho_{E}$ is the Larmor radius of protons with energy $E$. Such frequencies are indicated by the vertical dashed lines in Figure \ref{fig:psd15052005}.
The presence of a broad ${\rm PSD}(f)\propto f^{-1}$ range suggests that fluctuations are freshly-injected in that region and that turbulence has not become fully-developed yet. In addition, the presence of a bumped spectrum close upstream for the 2012 July 14 and 2003 November 4 crossings can be related to fluctuations driven by the shock-reflected energetic ions. 

Indeed, the PSD close upstream decays as a $f^{-1}$ power-law with a bend over towards frequencies higher than the ones corresponding to energetic particle scales (see Figs. \ref{fig:psd14072012} and \ref{fig:psd04112003}). Notice that in these two events, the power in the magnetic field fluctuations at low frequencies is smaller (or almost comparable) in the close upstream region than far upstream (see Figs. \ref{fig:psd14072012} and \ref{fig:psd04112003} and the discussion below). 

\begin{figure}[t]
    \centering
    \includegraphics[scale=0.6]{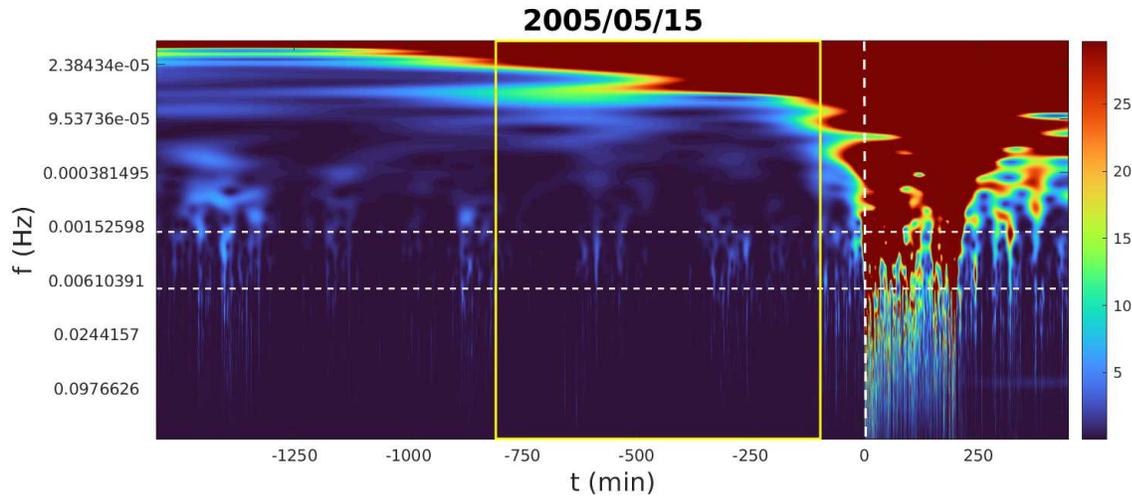}
    \caption{Power spectral density computed using the wavelet coefficients over the three magnetic field components (see text for further details). The yellow box delimits the region upstream where a flat energy spectrum has been detected, while the vertical dashed line remarks the shock crossing time. The horizontal dashed lines indicates the frequency range corresponding to the Larmor radius of energetic particles from $67$ keV to $2$ MeV. Typical cascade patches \citep{Greco16} can be recognized over the entire upstream region, with an intensity increase close to the shock front. Just behind the shock the CME sheath region is characterized by a very high level of turbulence over a broad range of frequencies.}
    \label{fig:psdwave15052005}
\end{figure}

Further, we have explored how turbulent fluctuations are distributed both in frequency and as a function of the distance from the shock via a wavelet analysis of the magnetic field vector. The square of the wavelet coefficients of the magnetic field components have been computed as \citep{Alexandrova08}
\begin{equation}
    |\mathcal{W_{\bf B}}(\tau,t)|^2 = \sum_i |\mathcal{W}_i(\tau,t)|^2,
    \label{eq:waveletenergy}
\end{equation}
where $\tau=1/f$ represents a time scale, and the sum is over the magnetic field components, $i=R, T, N$.
$\mathcal{W}_i(\tau,t)$ represent the Morlet wavelet coefficients computed over different $\tau$ and time $t$~\citep{Torrence98}, i.e., $\mathcal{W}_i(\tau,t) = \sum_{j=1}^{N} B_i(t_j)\psi^*\left[ \left( t_j-t \right)/\tau \right]$, with $\psi^*$ being the conjugate of the wavelet function. This allows us to assess the magnetic energy content in frequency and in time and localize, within the time series, the regions with high magnetic energy. Eq. (\ref{eq:waveletenergy}) has been reported for the 2005 May 15 shock crossing in Figure \ref{fig:psdwave15052005}.
It is indeed evident how the magnetic field fluctuation power increases close to the shock front, over a broad frequency range, which includes also the frequencies corresponding to the Larmor radius of the energetic particles, indicated in Figure \ref{fig:psdwave15052005} by the horizontal white dashed lines. The yellow box surrounds the flat energy spectrum region. Thus, the magnetic energy stored in the fluctuations tends to increase very close to the shock (indicated by the vertical dashed line) and is found to be high downstream of the shock within the turbulent sheath region. Here fluctuations are highly compressed and enhanced.
The increase of magnetic power close to the shock reconciles with the observation of an extended $\sim f^{-1}$ range in the PSD computed in the close upstream region, which is also in agreement with recent observations of turbulence close to IP shocks \citep{Zhao2021}. The local increase of the magnetic field power and the detection of a bumped PSD close upstream might be ascribed to a self-generated, freshly-injected turbulence, due to the presence of the energetic particle fluxes.

\begin{figure}[t]
    \centering
    \includegraphics[scale=0.6]{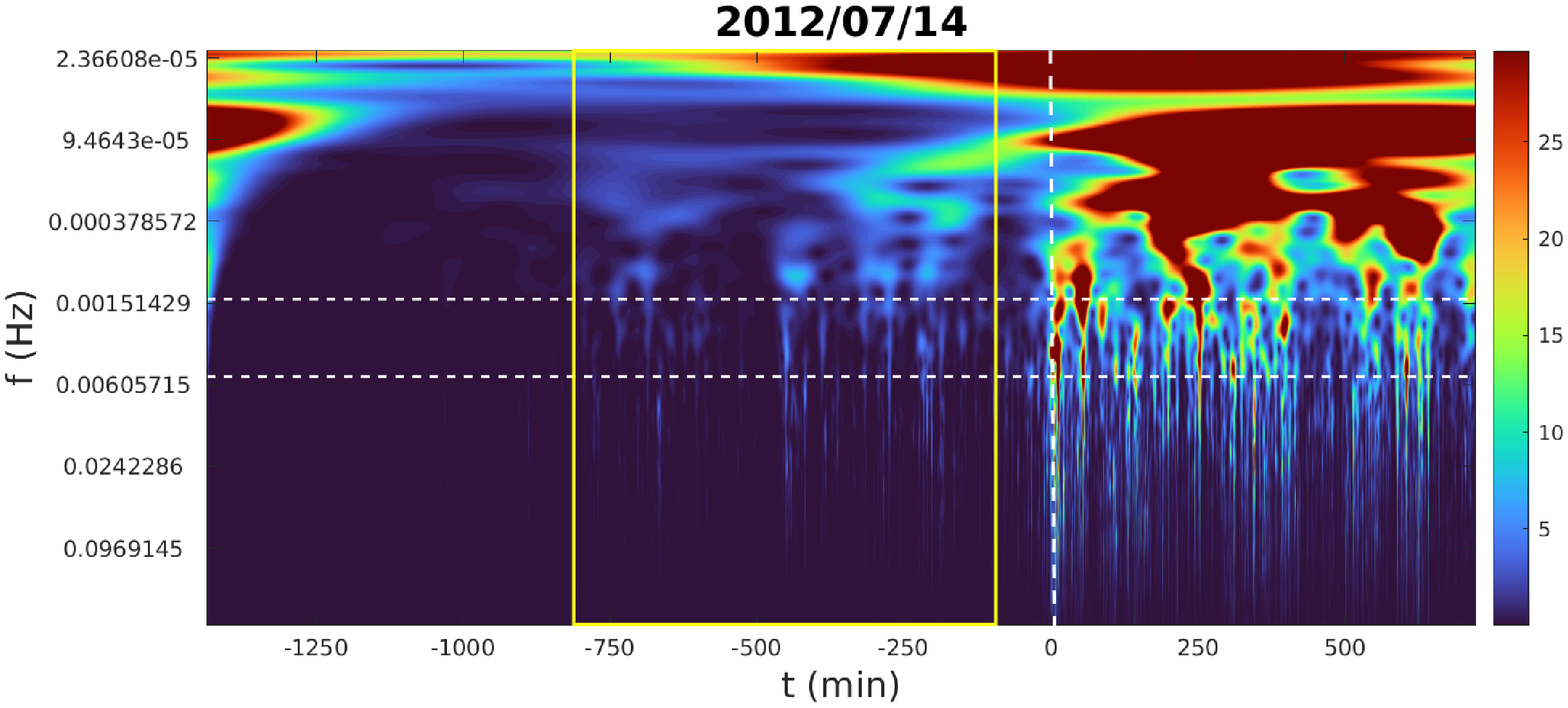}
    \caption{Same as Figure \ref{fig:psdwave15052005} but for the shock crossing by the ACE spacecraft on 2012 July 14.}
    \label{fig:psdwave14072012}
\end{figure}

In addition, from the scalograms in Figs. \ref{fig:psdwave14072012} and \ref{fig:psdwave04112003}, large amplitude pre-existing magnetic fluctuations can be observed over a broad range of frequencies and times. This evidence is in agreement with the comparable power found in the spectra upstream and close upstream in the 2012 July 14 and in the 2003 November 4 events. Such larger amplitude fluctuations far upstream can favour particle scattering, especially at lower frequencies where more power is stored. This can explain the tendency of high energy ion fluxes to separate from the fluxes in the other energy channels in the 2012 July 14 and in the 2003 November 4 events.

\subsection{Energetic Particle Anisotropy}
\begin{figure}
    \centering
    \includegraphics[scale=0.3]{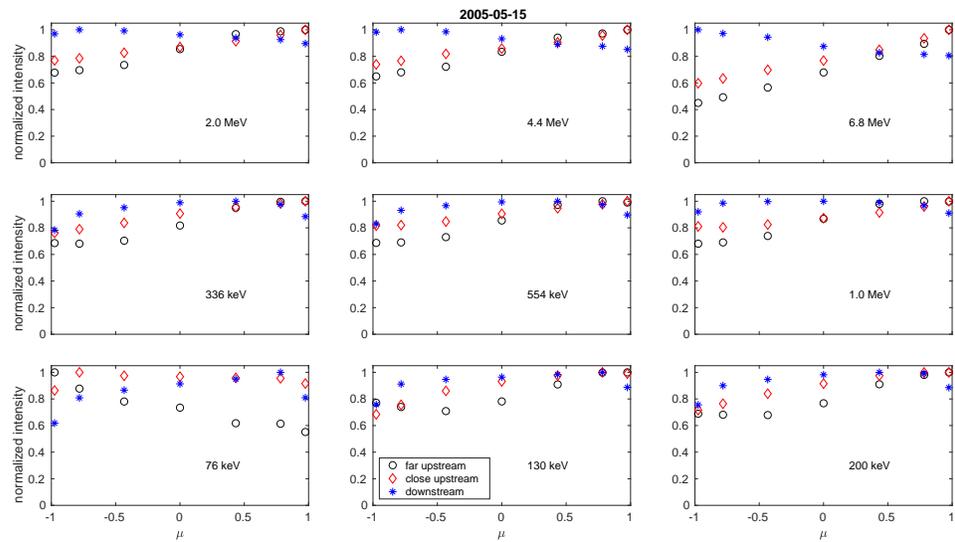}
    \caption{Mean ion fluxes, normalized to the maximum intensity in each region (see text for further details), as measured from the Wind/3DP/SST instrument in the solar wind frame within each pitch-angle bin during the 2005 May 15 event. Each panel refers to a given energy channel. The black circles indicate the ion fluxes measured far upstream of the shock wave within the shaded region indicated in Figure \ref{fig:15052005}, the red diamonds refer to the close upstream region, where ion fluxes start separating, and the blue stars to the downstream sheath region. 
    }
    \label{fig:fluxpitchangle}
\end{figure}

\begin{figure}[t]
    \centering
    \includegraphics[scale=0.3]{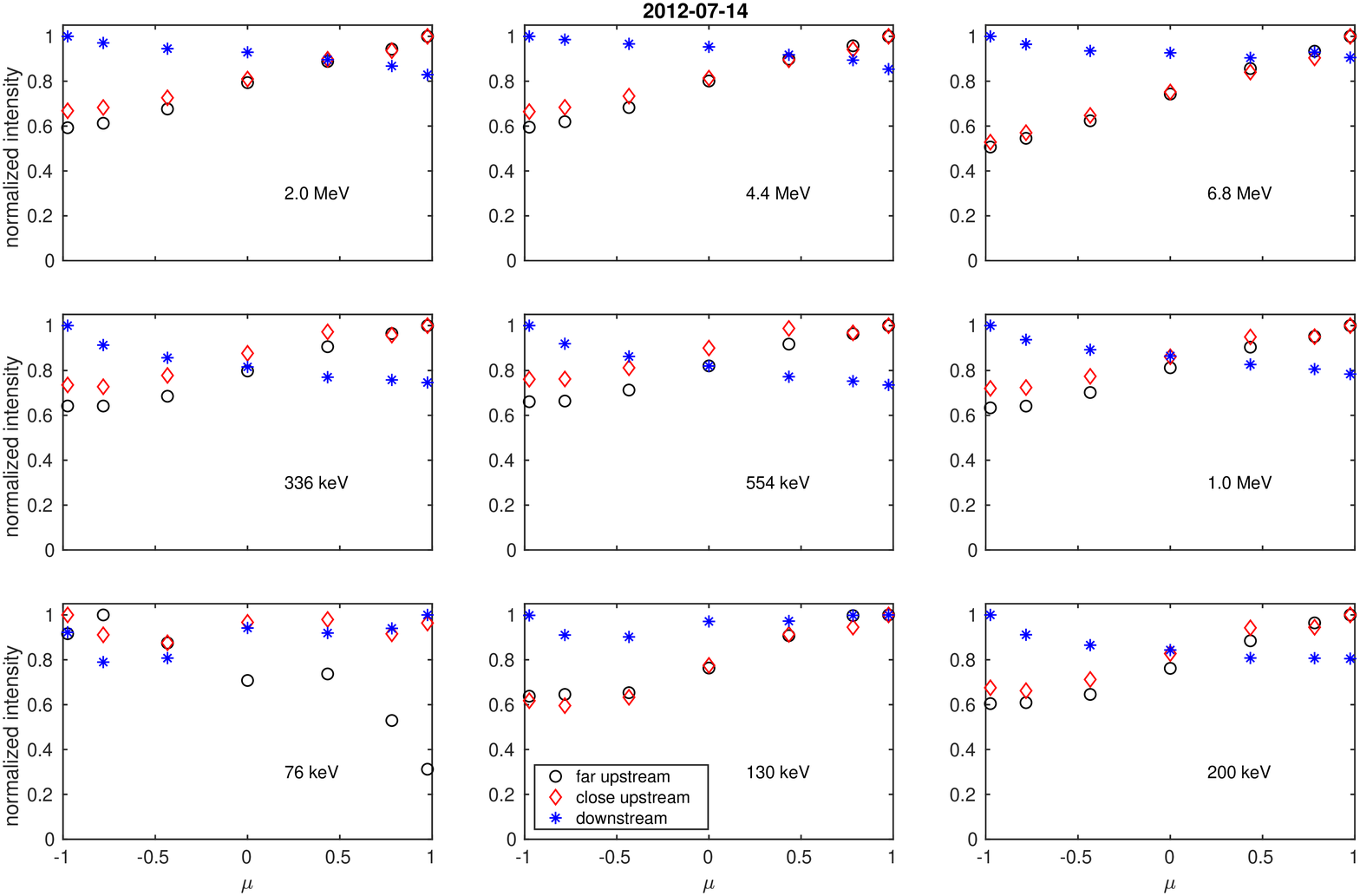}
    \caption{Same as Figure \ref{fig:fluxpitchangle} but for the shock crossing by the ACE spacecraft on 2012 July 14th.}
    \label{fig:fluxpitchangle2012}
\end{figure}

At this point is crucial to investigate the propagation of energetic particles with respect to the magnetic field direction using the Solid State Telescope (SST) of Wind 3DP \citep{Lin95}, in order to better characterize their motion. Ion fluxes are organized in $9$ energy channels with average energies of $76$ keV, $130$ keV, $200$ keV, $336$ keV, $554$ keV, $1.0$ MeV, $2.0$ MeV, $4.0$ MeV, $6.8$ MeV. They are also binned in $8$ pitch-angle values with respect to the local magnetic field direction; such fluxes, normalized to their maximum intensity, are displayed in nine panels in Figure \ref{fig:fluxpitchangle} for the 2005 May 15 event, as a function of the pitch-angle cosine $\mu$. Fluxes are in the solar wind frame, in order to get rid of any anisotropic feature due to the Compton-Getting effect \citep[][]{Compton35,Forman70}. The three sets of symbols in Fig. \ref{fig:fluxpitchangle} refer to the mean fluxes calculated over three different regions around the shock: far upstream where flat spectra are detected (from -750 min to -200 min upstream of the shock), close upstream of the shock (from -80 min to -10 min), and from 10 min up to 70 min downstream of the shock. In all energy channels there is a tendency to isotropize the distributions of the fluxes in the downstream region, caused by the presence of the augmented magnetic field fluctuations which efficiently scatter energetic particles. Far upstream, within the region of overlapped fluxes, we observed for the lowest energy channel that the flux is higher at large pitch-angles, namely particles move sunward towards the shock, while higher energy particles travel mostly anti-sunward at small pitch-angles. Here the amplitude of magnetic field fluctuation decreases but the field is almost radial and a good connection between the shock and the spacecraft exists. This permits to detect in the far upstream region those particles that have been isotropized close upstream, considering that parallel diffusion is larger than perpendicular diffusion. This anisotropy in pitch angle is somewhat reduced for all the energy channels in the close upstream region. This is again due to the amplified magnetic fluctuations, since ions can easily be scattered in all directions by interacting with turbulent fluctuations \citep[][]{Giacalone1999,Trotta21}. Here, lower energy particles tend to be isotropized in $\mu$, while there is still anisotropy in favour of field-aligned particles at higher energies. A similar behaviour has been observed for the energetic particle fluxes associated to the 2012 July 14 event (see Fig. \ref{fig:fluxpitchangle2012}).


\begin{figure}[t]
    \centering
    \includegraphics[scale=0.6]{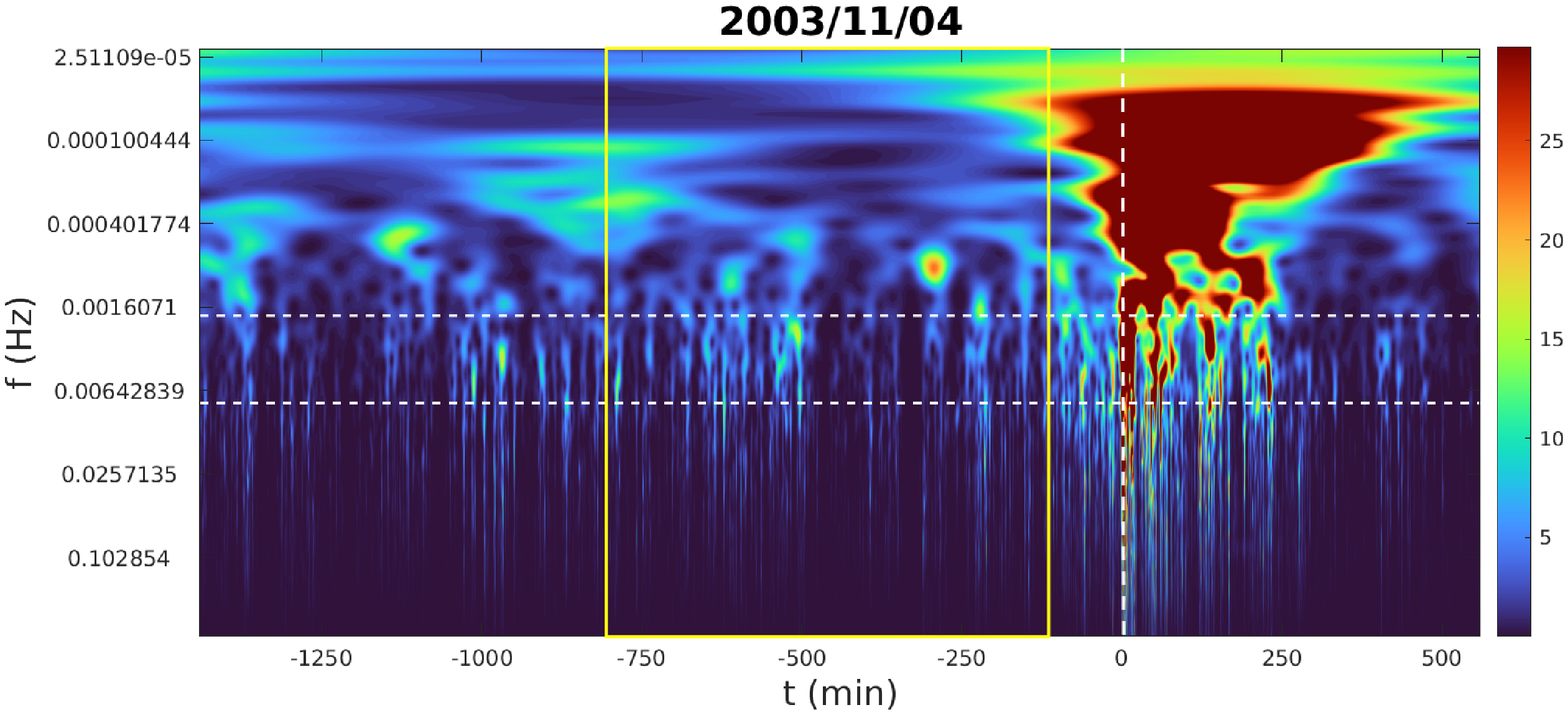}
    \caption{Same as Figure \ref{fig:psdwave15052005} but for the shock crossing by the ACE spacecraft on 2003 November 4.}
    \label{fig:psdwave04112003}
\end{figure}

\begin{figure}[t]
    \centering
    \includegraphics[scale=0.3]{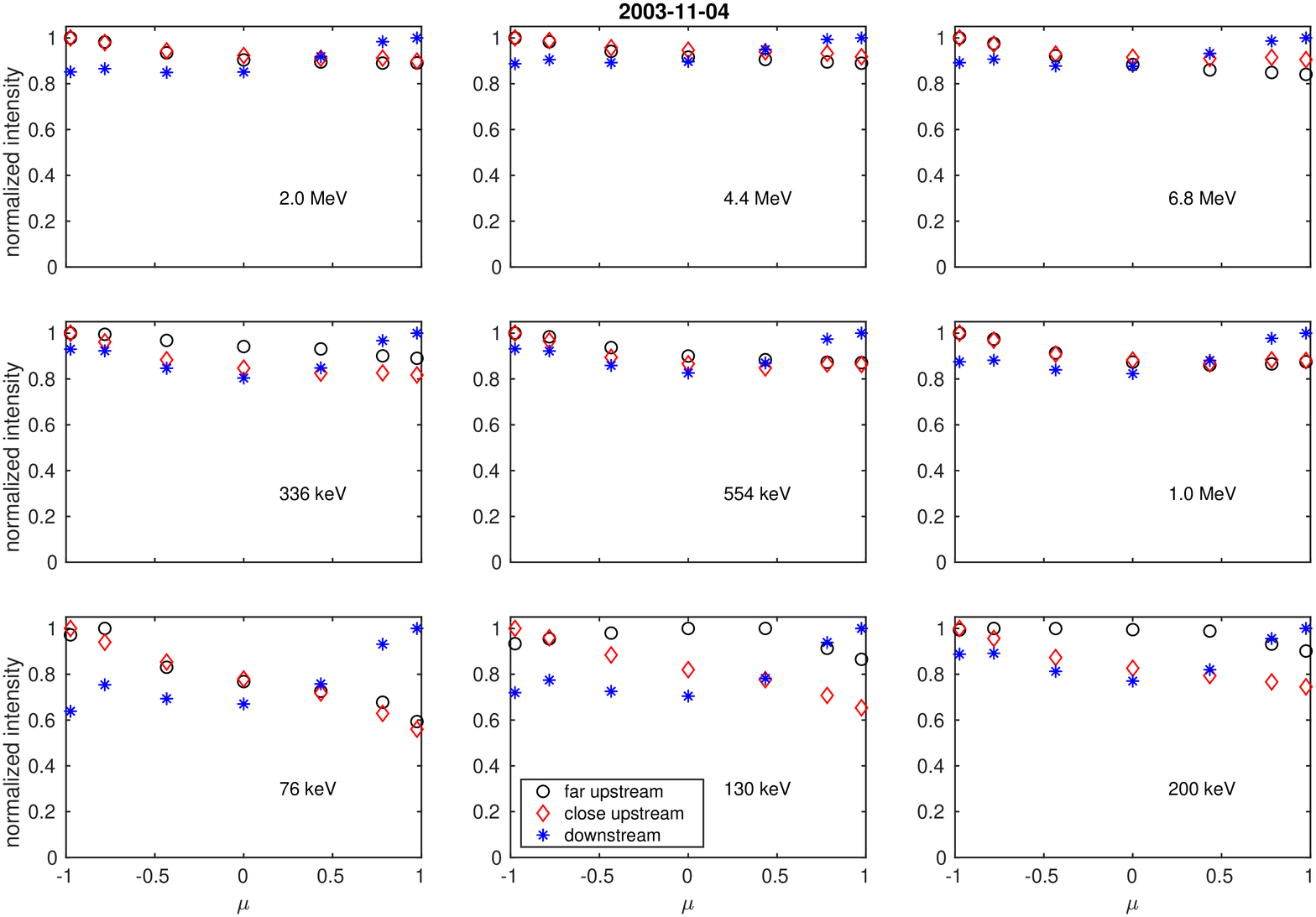}
    \caption{Same as Figure \ref{fig:fluxpitchangle} but for the shock crossing by the ACE spacecraft on 2003 November 4.}
    \label{fig:fluxpitchangle2003}
\end{figure}

On the other hand, the mean ion flux values displayed in Fig. \ref{fig:fluxpitchangle2003} for the 2003 November 4 shock exhibit a high degree of isotropy in the three regions for almost all the energy channels, except for the $76$ keV channel, where particles are almost anti-aligned to the mean field (which is sunward), going along the anti-sunward direction. Notice from Fig. \ref{fig:04112003} that the flux of energetic ions in this channel is not completely overlapped to the other fluxes in the shaded region. Observing Fig. \ref{fig:psd04112003} and Fig. \ref{fig:psdwave04112003}, the power stored in the magnetic field fluctuations is a bit higher than in the other two events. This will induce more particle scattering and therefore more isotropy.

\section{Discussion on the velocity filter condition}
\label{sec:discussion}

\begin{figure}
    \centering
    \includegraphics[scale=0.4]{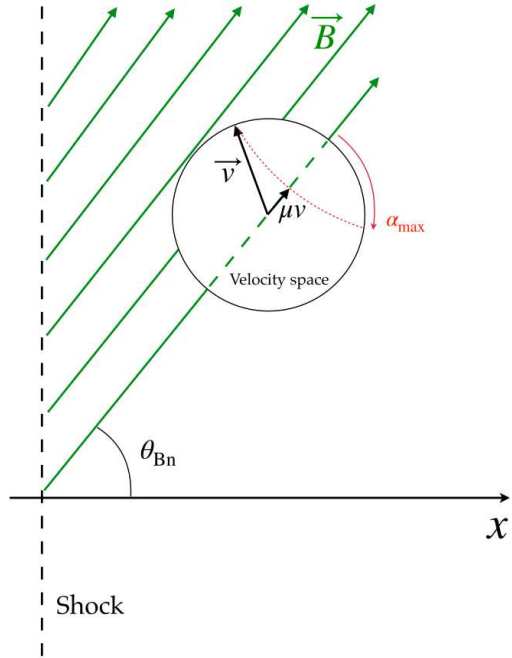}
    \caption{Cartoon of the spherical sector around the magnetic field direction with radius given by the particle velocity $\mathbf{v}$.}
    \label{fig:shell}
\end{figure}

Following the interpretation proposed by \citet{LeRoux12} \citep[see also][]{lario19}, we have derived the energetic particle flux by imposing the velocity filter condition, namely low energy particles cannot easily escape far upstream. Since the number of particles within a volume in phase space is defined as \citep[e.g.,][]{Moraal13}
\begin{equation}
    d{\cal N}=F(\mathbf{r},\mathbf{p},t) d^3r d^3p=F(\mathbf{r},\mathbf{p},t) d^3r p^2 dp d\Omega,
    \label{eq:N}
\end{equation}
where $F(\mathbf{r},\mathbf{p},t)$ is the particle distribution function and $d\Omega=\sin{\alpha} d\alpha d\phi$ is the element of solid angle in momentum space, we can define a differential density in momentum, namely the number of particles in a given volume $d^3r$ and with momentum within $\mathbf{p}$ and $\mathbf{p}+d\mathbf{p}$ as 

\begin{equation}
    U_p=p^2 \int_{\Omega} F(\mathbf{r},\mathbf{p},t) d\Omega.
    \label{eq:Up}
\end{equation}
This implies that the number of particles becomes $dN=U_p d^3r dp$. We align the polar coordinate axis with the upstream magnetic field, so that $\alpha$ is the pitch angle; considering that only particles with a sufficiently large parallel velocity can escape from the moving shock, we find that $\alpha$ can vary over a limited range of values, namely the particle velocity belongs to a limited spherical sector in phase space (see the cartoon in Figure \ref{fig:shell}). Then, we can calculate the  differential density of \textit{upstream propagating particles} as 
\begin{equation}
    U'_p=2\pi p^2 \int_0^{\alpha_{max}} F(\mathbf{r},\mathbf{p},t) \sin{\alpha} d\alpha.
    \label{eq:Up_modified}
\end{equation}
Here, we assume that $F(\mathbf{r},\mathbf{p},t)$ is isotropic, i.e., that the particle distribution function depends on the momentum magnitude $p$ only. 
On the other hand, the velocity filter involves a condition on $\mu$ and therefore generates anisotropic distributions, as those which are commonly observed far upstream, see for examples Fig.s \ref{fig:fluxpitchangle}-\ref{fig:fluxpitchangle2003}. In other words, near isotropy is assumed close upstream of the shock but the escaping particles will have an anisotropic distribution function.
We also notice that the effect of the velocity filter is easily quantified assuming an isotropic $F$; conversely, if some dependence on $\mu$ would be given, the integration in Eq. (\ref{eq:Up_modified}) would be less straightforward but a velocity filter would still be acting. With the isotropy assumption, we can insert the omnidirectional distribution function $f(\mathbf{r},p,t)$ in Eq.(\ref{eq:Up_modified}); using the pitch-angle cosine $\mu = \cos\alpha$, defining a minimum pitch-angle cosine $\mu_{min}$ corresponding to $\alpha_{max}$, and being $d\mu=-\sin{\alpha} d\alpha$, one can readily find that 
\begin{equation}
    U'_p=2\pi p^2 f(\mathbf{r},p,t) (1-\mu_{min}) = 
    2\pi p^2 f(\mathbf{r},p,t) \biggr(1-\frac{V_1^{sh}\sec{\theta_{\rm Bn}}}{v}\biggl).
    \label{eq:Up_final}
\end{equation}
Eq.(\ref{eq:Up_final}) introduces a correction to the isotropic (in pitch-angle) flux of particles and the correcting factor comes from the condition for particle escape from the shock $\mu v> V_1^{sh}\sec{\theta_{\rm Bn}}$ \citep{lario19}. Thus, the differential density is related to the flux $J=U'_p\propto p^2f(p)$ \citep{Moraal13}; according to the DSA prediction, we have $f(p)\propto p^{-3r/(r-1)}$, where $r=n_2/n_1$ is the compression ratio of the shock.  

\begin{figure}[h]
    \centering
    \includegraphics[scale=0.23]{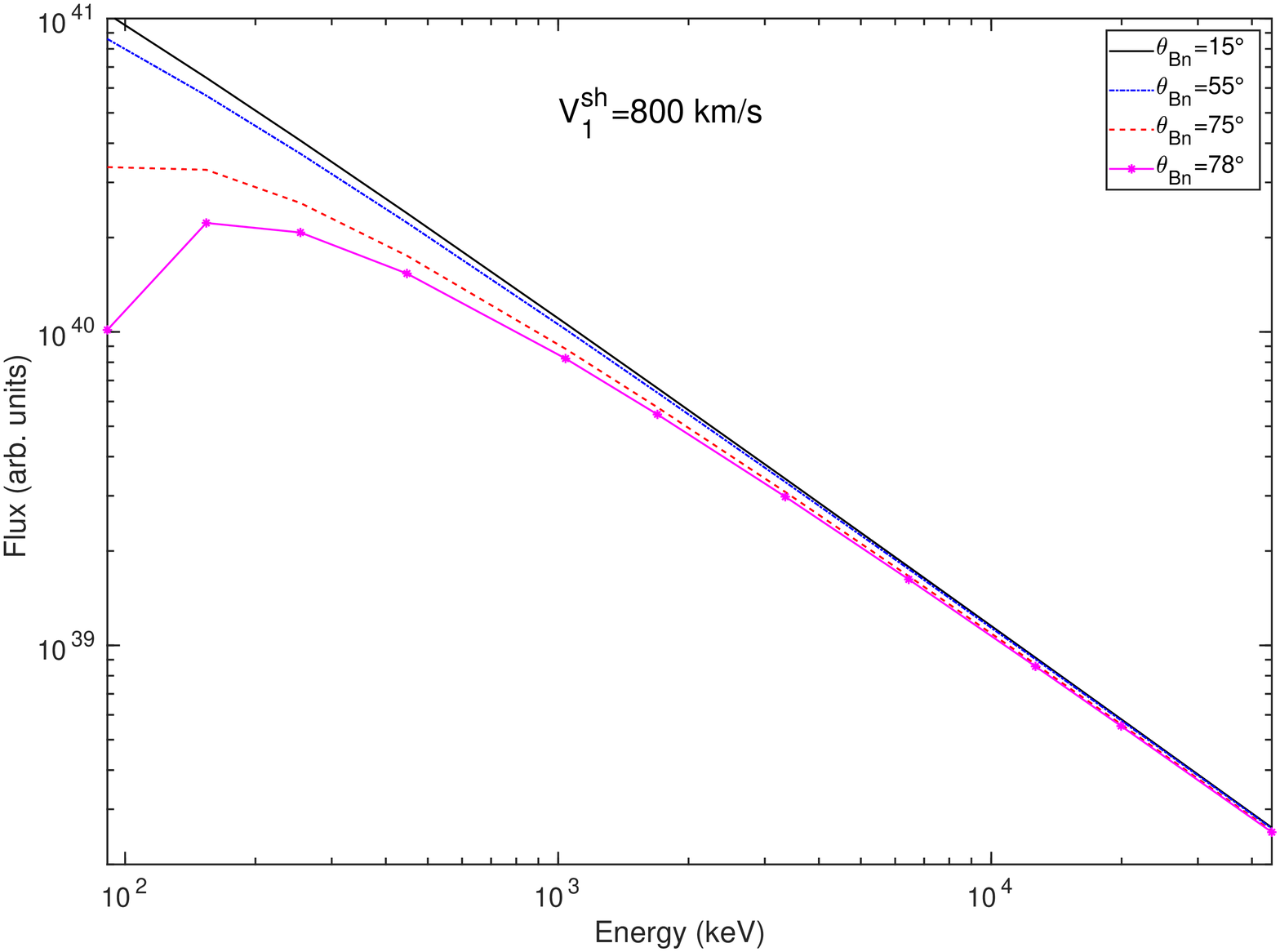}
    \includegraphics[scale=0.26]{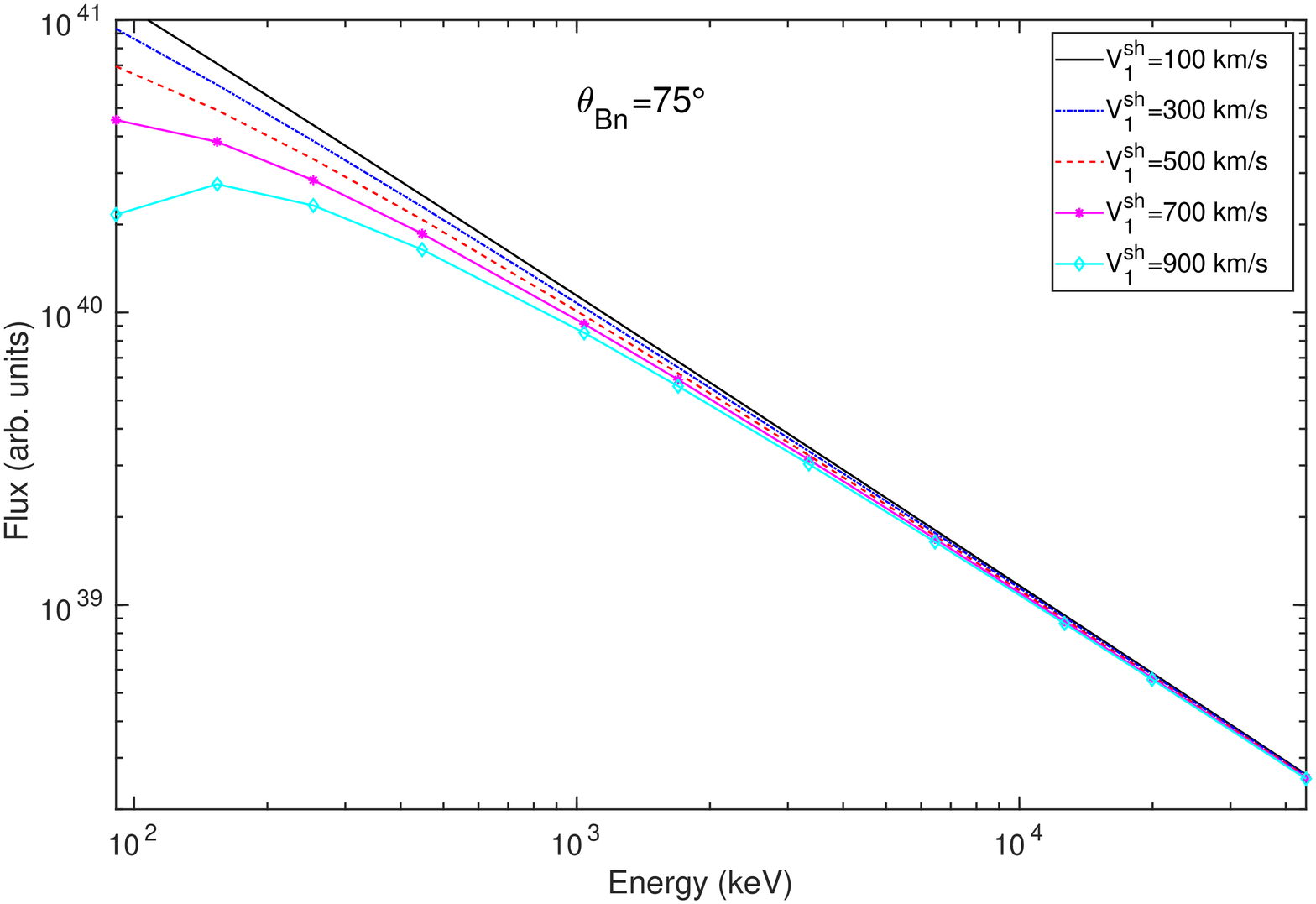}
    \caption{Left panel: Particle energy spectrum computed fixing the upstream plasma speed in the shock reference frame at $V_1^{sh}=800$ km/s and by varying the $\theta_{\rm Bn}$ angle. The velocity filter acts for quasi-perpendicular shocks at low energies, causing a flattening in the spectrum. Right panel: Particle energy spectrum computed fixing $\theta_{\rm Bn}=75^{\circ}$ and by varying $V_1^{sh}$. The velocity filter acts for high speed plasma flows at low energies, causing a flattening in the spectrum.}
    \label{fig:spectrum_thetabn}
\end{figure}

In order to understand the effect of such a velocity filter on the particle energy spectrum observed upstream of shocks, we have computed the flux $J$ over a broad energy range (from about $70$ keV to $40$ MeV) using Eq.(\ref{eq:Up_final}). First, we have fixed the plasma speed in the shock rest frame and have varied $\theta_{\rm Bn}$  from quasi-parallel to quasi-perpendicular values. The left panel in Figure \ref{fig:spectrum_thetabn} shows a flattening at low energies for quasi-perpendicular shocks. This means that the flux of those particles is substantially reduced under those conditions and this promotes the observations of flat spectra upstream of the shocks. Notice that the observed reduction of the flux involves energy channels below about $400$-$600$ keV, which are the channels where the upstream energetic particle fluxes are frequently seen to be overlapped 
\citep[see also][for the $\theta_{\rm Bn}$ dependence]{Zim20}. 

We have further plotted the energy spectrum by fixing $\theta_{\rm Bn}=75^{\circ}$ and by varying the upstream plasma speed in the shock frame of reference. Thus, the right panel in Figure \ref{fig:spectrum_thetabn} displays a reduction of the low energy flux for high values of the plasma speed. This is because the particle escape condition tends to be marginally fulfilled for high $V_1^{sh}$ values.

\begin{figure}
    \centering
    \includegraphics[scale=0.4]{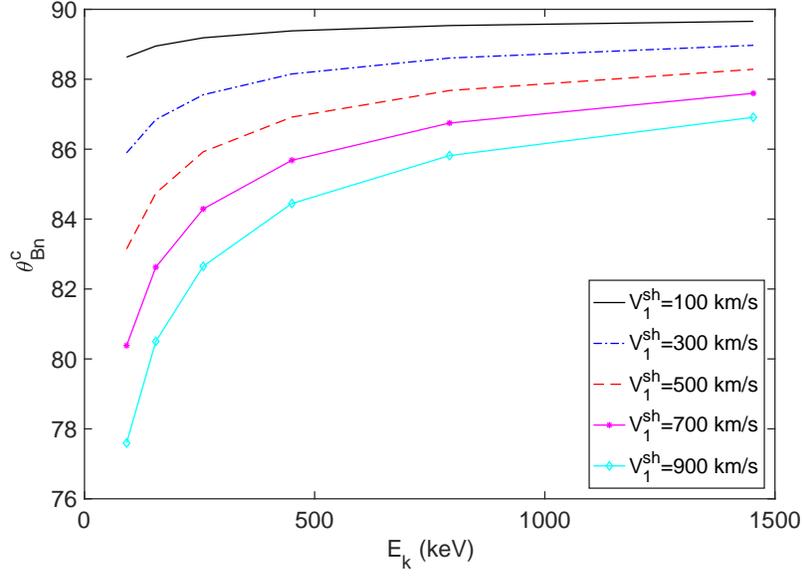}
    \caption{$\theta^c_{\rm Bn}$ as a function of particle energy for several values of the upstream plasma speed (in the shock frame).}
    \label{fig:thetabn}
\end{figure}

Further, from the above condition it is possible to derive a limit value of $\theta_{\rm Bn}$ for particle upstream propagation. Namely, since the cosine of the pitch-angle is a bounded quantity, i.e., $-1\le\mu\le 1$, the particle escape condition when $|\mu|=1$ becomes $v=V_1^{sh}\sec{\theta^c_{\rm Bn}}$, so that a critical value for the angle between the magnetic field direction and the normal-to-the-shock direction can be easily found,
\begin{equation}
    \theta^c_{\rm Bn}=\arccos \biggr(\frac{V_1^{sh}}{v}\biggl).
    \label{eq:thetabnc}
\end{equation}
Figure \ref{fig:thetabn} displays the critical $\theta^c_{\rm Bn}$ as a function of particle energy for several values of the plasma speed in the shock frame: when $\theta_{\rm Bn}>\theta^c_{\rm Bn}$ within a given energy range the particle escape condition cannot be  satisfied and particles can be either confined close to the shock front or transmitted downstream. As $V_1^{\rm sh}$ increases, the escape condition is marginally satisfied, especially at low ion energies.

\section{Conclusions}
\label{sec:conclusions}

In this paper we have reported peculiar observations of flat particle energy spectra upstream of three interplanetary shock waves, as detected by ACE and Wind. In all of these events the spacecraft was well connected to the shock front, so that it has been possible to observe fluxes of particles previously accelerated at the shock. The analysis performed on the magnetic field measurements, by means of the computation of the PSD of the field components, has revealed that close upstream the PSD at the time scales in resonance with energetic particles tends to be bumped, and then bends over when going towards smaller time scales. This suggests the presence of newly injected magnetic field fluctuations, probably due to the presence of streaming energetic particles. This, of course, favours particle scattering and diffusion near the shock front. Furthermore, Wind/3DP/SST observations of energetic ion fluxes in different pitch-angle bins have highlighted isotropic (in pitch-angle) distributions downstream of the shocks and isotropy is also recovered for lower energy particles close upstream, where magnetic fluctuations are enhanced. Higher energy particles tend to be more anisotropic close upstream and far upstream, suggesting that they can escape more easily from the region in which lower energy particles are confined, thus promoting the formation of a flat energy spectrum. This scenario also supports the numerical results by \citet{Ng03}. 

We have explained these observations by elaborating the suggestion by \citet{lario19} that a velocity filter, depending on both the particle speed and the pitch-angle, favours the upstream propagation of faster particles, leading to the  flattening of the energy spectrum and the  overlapping of the fluxes for moderate energies. We have derived the differential density of upstream propagating particles and we have found that for a given $\theta_{\rm Bn}$, particles with higher energies/velocities and large $\mu$ tend to fulfill the velocity filter condition and can be easily detected far upstream, while lower energy particles with small $\mu$ cannot match that condition and tend to be confined at the shock front. As a consequence, their fluxes are depleted far upstream and this leads to flat spectra. Such a mechanism would lead to particle anisotropy, and the degree of anisotropy should depend on both the velocity filter and the scattering undergone by energetic particles. 
In addition, the energetic particle fluxes have been computed both as a function of the shock normal angle $\theta_{\rm Bn}$ and as a function of the shock speed. It is found that the effect of the velocity filter is the more relevant, the more perpendicular the shock is. Such a flattening compares well with the flux observations of the three shocks here analyzed, since the energy channels where a significant depletion of the flux is observed match the ones in which overlapped ion fluxes are detected in the observations (i.e., from $70$ keV up to $\sim 600$ keV). 
On the other hand, these flat spectra are observed several hours prior to the shock arrival, with particles being accelerated by a shock having different properties than those of the shock crossing, and in particular different shock normal angles $\theta_{\rm Bn}$. In this regard, we can consider that for a CME propagating not too far from the Sun-Earth line, and because of the typical shape of the Parker spiral magnetic field, spacecraft at L1 are connected with the westward part of the CME driven shock, which can be considered to be more ``perpendicular" than the shock near to Sun-Earth line. This allows for an efficient active influence of the velocity filter effect. It is also interesting to note that the spectral flattening is maintained for a large range of distances between the approaching shock and Earth, something which suggests that the energetic particle transport properties might be nearly independent of the particle energy.

In summary, here we have reported and interpreted the observational feature of flat energy spectra upstream of some interplanetary shocks by using magnetic field, energetic particles, and plasma measurements. This joint analysis has allowed us to devise the following scenario: close to the shock front magnetic field fluctuations are injected at the scales of energetic particles, this creates an environment in which lower energy particles can be efficiently scattered. Indeed, the pitch-angle distributions for those particles have been found to be almost isotropic. Higher energy particles can easily go far upstream outward from the shock because of the velocity filter condition. For lower energy particles, the escaping condition is not satisfied in case of fast, quasi-perpendicular shocks (as it is almost the case for the three shocks analyzed). Their flux is then depleted and this promotes the formation of a flat energy spectrum far upstream.
In future works we foresee to support such a scenario with Solar Orbiter and Parker Solar Probe observations closer to the Sun and to investigate more deeply the role of the shock geometry on the velocity filter condition from in-situ measurements.

D.L. acknowledges support from NASA Living With a Star (LWS) programs NNH17ZDA001N-LWS and NNH19ZDA001N-LWS, the Goddard Space Flight Center Internal Scientist Funding Model (competitive work package) program, and the Heliophysics Innovation Fund (HIF) program.
The work was supported by the Geospace Environment Modeling (GEM) Focus Group "Particle Heating and Thermalization in Collisionless Shocks in the MMS Era" led by L.B. Wilson III.



\bibliography{big_bib} 






\end{document}